\documentclass[twocolumn,showpacs,preprintnumbers,amsmath,amssymb,prl]{revtex4} \usepackage{graphicx}
\usepackage{dcolumn}
\usepackage{color}
\usepackage{bm}
\begin{document} 
%%%%%%%%%%%%%%%%%%%%%%%%%%%%%%%%%%%%%%%%%%%%%%%%%%%%%%%%%%%%%%%%%%%%%%%%%%%%%%%%
\newcommand{\ket}[1]{|#1\rangle}
\newcommand{\bra}[1]{\langle#1|}
\newcommand{\braket}[2]{\langle#1|#2\rangle}
\newcommand{\kb}[2]{|#1\rangle\langle#2|}
\newcommand{\kbs}[3]{|#1\rangle_{#3}\phantom{i}_{#3}\langle#2|}
\newcommand{\kets}[2]{|#1\rangle_{#2}}
\newcommand{\bras}[2]{\phantom{i}_{#2}\langle#1|}
\newcommand{\af}{\alpha}
\newcommand{\bt}{\beta}
\newcommand{\gm}{\gamma}
\newcommand{\la}{\lambda}
\newcommand{\dt}{\delta}
\newcommand{\s}{\sigma}
\newcommand{\qq}{(\s_{y}\otimes\s_{y})}
\newcommand{\uu}{\rho_{12}\qq\rho_{12}^{*}\qq}
\newcommand{\tr}{\textrm{Tr}} 
%%%%%%%%%%%%%%%%%%%%%%%%%%%%%%%%%%%%%%%%%%%%%%%%%%%%%%%%%%%%%%%%%%%%%%%%%%%%%%%%
%%%%%%%%%%%%%%%%%%%%%%%%%%%%%%%%%%%%%%%%%%%%%%%%%%%%%%%%%%%%%%%%%%%%%%%%%%%%%%%%
\title {\bf Environment assisted entanglement enhancement}
\author{Biplab Ghosh}
%\altaffiliation{biplab@bose.res.in}
\affiliation{S. N. Bose National Centre for Basic Sciences,
Salt Lake, Kolkata 700 098, India}
\author{A. S. Majumdar}
\altaffiliation{Corresponding author email:archan@bose.res.in}
\affiliation{S. N. Bose National Centre for Basic Sciences,
Salt Lake, Kolkata 700 098, India}
\author{N. Nayak}
%\altaffiliation{nayak@bose.res.in}
\affiliation{S. N. Bose National Centre for Basic Sciences,
Salt Lake, Kolkata 700 098, India}
\date{\today}

\vskip 0.5cm                              
\begin{abstract}
We consider various dissipative atom-cavity systems and show that their 
collective dynamics can be used to maximize entanglement for intermediate 
values of the cavity leakage parameter $\kappa$. We 
first consider the interaction of a single two-level atom with one of two
coupled microwave cavities and show analytically that the atom-cavity 
entanglement increases with cavity leakage. We next consider
a system of two atoms passing successively through a cavity and derive the 
expression for the maximum value of $\kappa$ in terms of the Rabi angle $gt$, 
for which the two-atom entanglement can be increased. Finally, numerical 
investigation of micromaser dynamics also reveals the increase of two-atom 
entanglement with stronger cavity-environment coupling for experimentally 
attainable values of the micromaser parameters. 
\end{abstract}                                                                 
\pacs{03.67.Mn, 03.65.Yz, 42.50.Pq}
\maketitle

Quantum entanglement has emerged as a highly useful resource in communication
and computation protocols\cite{Nielsen} in recent years. There have been 
many proposals for generating atomic entanglement and entanglement 
between cavity
modes through atom-photon interactions\cite{proposals}, and
some notable experimental
demonstrations have also been performed\cite{examples}.
Several interesting
features and applications of entanglement can be obtained in 
devices involving optical and microwave
cavities\cite{Raimond}. In all these schemes interaction with the surrounding
heat bath and cavity leakage has to be monitored such that rapid 
decoherence\cite{zurek} is unable to destroy the created entanglement
within the time-frame required for observation. 

Though most proposals of entanglement generation rely on methods to reduce
the coupling with the environment, there have been some 
suggestions\cite{plenio,braun,knight,others,plenio2}
for creating entanglement between two or more parties by their collective 
interactions with a common environment. The effectiveness of collective 
interactions
in the dynamics of quantum optical systems has been appreciated
much earlier\cite{book}. 
Specific examples of environment induced entanglement
have been worked out recently. Schemes of using the decay of the cavity field
to induce atom-atom or cavity-cavity entanglement have been 
proposed\cite{plenio}. Braun has shown that entanglement can be created 
between two qubits which do not interact directly with each other, but
interact with a common heat bath\cite{braun}. A corollary of this result
is the mediation of atomic entanglement by a thermal
field inside a cavity\cite{knight}. Other proposals involving the collective
dynamics of trapped ions, squeezed and thermal fields, and quantum-reservoir
engineering have also been suggested\cite{others,plenio2}.
The purpose of the present work is to show not just the creation
of entanglement with environmental assistance, but to demonstrate the 
feasibility of actual enhancement of entanglement in real workable devices.  

To this end in this Letter we show that entanglement in atom-cavity 
devices can be quantitatively increased by increasing the cavity damping
rate. We illustrate our point by first considering two examples of tripartite
systems ((i)two cavities and a single atom, and (ii) one cavity and two atoms)
where we obtain analytically the expressions for the atom-cavity and two-atom
concurrences, respectively, as functions of the cavity leakage parameter
$\kappa$. It is seen explicitly that the concurrences are maximised for
intermediate values of $\kappa/g$ (where $g$ is the atom-cavity coupling 
constant). 
Plenio and Hulega\cite{plenio2} have earlier shown numerically
that entanglement between two optical cavities driven by an external 
optical white noise field can be maximised for intermediate values of the 
cavity damping rates. In this work we derive analytically a similar result, 
using simple but practically realizable systems with two examples.

We next consider the experimentally workable micromaser\cite{micro1}.
The micromaser device is well known for its utility in the generation
of entangled atomic states\cite{micro2}. The controlled
monitoring of dissipative effects makes it possible to study fundamental
aspects like nonlocality and information transfer through the 
micromaser\cite{micro3}. Many of these features have been demonstrated
in several experiments performed using the micromaser\cite{micro1}.
Here we choose certain experimentally achieved range of values for the
micromaser parameters\cite{micro1} and show through numerical analysis that 
the entanglement between
a pair of atoms can be increased with the increase of cavity 
damping $\kappa$ up to a certain range of its values.

\section*{A. A two-level atom interacting with one of two maximally entangled cavities}

We first consider two initially maximally entangled single-mode 
cavities ($C_1$ and $C_2$).  [Such a system can be prepared 
by sending a single circular Rydberg atom in its exited state through 
two identical and initially empty high-Q microwave cavities\cite{David}.] 
A two-level Rydberg atom $A_1$ prepared in the ground state
$\ket{g}$ passes through the cavity $C_1$. 
The resonant interaction between the two-level atom
and cavity mode frequency takes place with the Rabi angle $gt$. 
In presence of the cavity 
dissipation the dynamics of the flight of the atom is governed by the 
evolution equation
\begin{eqnarray}
\dot\rho=\dot\rho|_{\textrm{atom-field}}+\dot\rho|_{\textrm{field-reservoir}},
\label{b10}
\end{eqnarray} 
At temperature $T=0$ $K$  the average thermal photon number 
is zero. Since we are working with a two-level Rydberg atom, its lifetime 
is much larger compared to the atom-cavity interaction time and hence we can 
neglect the atomic dissipation.	The first term on the r.h.s. of Eq.(\ref{b10})
evolves under the usual Jaynes-Cummings interaction, and the second term
is given by
\begin{eqnarray}
\dot\rho|_{\textrm{field-reservoir}}=-\kappa(a^\dagger a \rho-2a\rho a^\dagger+
\rho a^\dagger a).
\label{b11}
\end{eqnarray} 
where $\kappa$ is the cavity leakage constant. 
We consider the approximation of a two-level cavity, i.e., the probability of 
getting a two-- or more than 
two--photon number state of the cavities 
is zero. Under the secular approximation\cite{secular} the time-evolved 
density matrix of the reduced state of the first cavity and the atom
($\rho(t)_{C_1C_2A_1}$) is given by
\begin{eqnarray}  
\rho(t)_{C_1A_1}=\textrm{Tr}_{C_2}(\rho_{C_1C_2A_1})
= \left(\begin{array}{ccccc}
\alpha_1 & 0 & 0 & 0\\
0 & \alpha_3 & -\alpha_4 & 0\\
0 & \alpha_4 & \alpha_2 & 0\\
0 & 0 & 0 & 0  \end{array}\right)
\label{b13}
\end{eqnarray}
in the basis $|0_1g_1>$,
$|0_1e_1>$, $|1_1g_1>$, and $|1_1e_1>$,  where $|0>$ and $|1>$ represent 
the $0$-photon and the $1$-photon number 
basis states of the cavities, respectively, and $\alpha_i$ are given by
\begin{eqnarray}
\alpha_1&=&(1-\frac{e^{-\kappa_1 t}}{2})e^{-2\kappa_2 t},\nonumber\\
\alpha_2&=&(\cos^2{gt})e^{-\kappa_1 t}(1-\frac{e^{-2\kappa_2 t}}{2}),\nonumber\\
\alpha_3&=&(\sin^2{gt})e^{-\kappa_1 t}(1-\frac{e^{-2\kappa_2 t}}{2}),\nonumber\\
\alpha_4&=&i(\sin{2gt})e^{-\kappa_1 t}(1-\frac{e^{-2\kappa_2 t}}{2}),
\label{b131}
\end{eqnarray}
where $\kappa_1$ and $\kappa_2$ are the cavity
leakage constants of $C_1$ and $C_2$, respectively. 

We quantify the entanglement using the measure {\it concurrence}, which
for a general state $\rho_{12}$ is defined as\cite{con}
${\it C}(\rho_{12})=\max(0, \sqrt\la_1-\sqrt\la_2-\sqrt\la_3-\sqrt\la_4)$, 
where the $\la_{i}$ are the  eigenvalues of $\uu$ in descending order.
We compute the  concurrence for $\rho(t)_{C_1A_1}$, which is given by
\begin{eqnarray}
{\it C}(\rho(t)_{C_1A_1})&=&2 \cos{gt}\sin{gt}e^{-\kappa_1 t}(1-\frac{e^{-2\kappa_2 t}}{2})
\label{b14}
\end{eqnarray}
If we take $\kappa_1=\kappa_2=\kappa$, one gets
\begin{eqnarray}
{\it C}(\rho(t)_{C_1A_1}= 2 {\it C}_{ideal}(e^{-\kappa t}-e^{-3\kappa t}/2)
\label{b141}
\end{eqnarray}
For small $\kappa$,  we can write
\begin{eqnarray}
{\it C}(\rho(t)_{C_1A_1} \thickapprox {\it C}_{ideal}(1+\kappa t).
\label{b15}
\end{eqnarray}
where ${\it C}_{ideal} = \cos{gt}\sin{gt}$ is the value of the concurrence
in the case of ideal cavities with no dissipation.
It  follows from Eq.(\ref{b14}) that
for large $\kappa$ one gets
${\it C}(\rho(t)_{C_1A_1} \to 0$, as expected.
This feature of dissipation assisted increase of entanglement is
observed for a range of values of $\kappa$, and the maximum of concurrence
is obtained for the value of  $\kappa/g=[{\mathrm ln}(3/2)]/(2gt)$ at 
fixed $gt$.
As an aside, it is interesting to note 
that the
entanglement between the two cavities $C_1$ and $C_2$ falls off with
increasing $\kappa$, thus providing a manifestation of the monogamous
nature of entanglement between the pairs $C_1A_1$ and $C_1C_2$. A monogamy
inequality\cite{ckw} for this situation has been verified\cite{monogam2}.
The increase of entanglement with $\kappa$ as seen in Eq.(\ref{b15}) follows
from the collective nature of the dynamics of the two cavities, as is
apparent from the structure of the elements of the atom-cavity state
given in Eqs.(\ref{b131}) where the $\alpha_i$'s are the sums of two
terms involving $\kappa_1$ and $\kappa_2$ respectively.  This motivates
us to look for similar collective effects in other simple tripartite systems 
such as the one involving
the interaction of a single cavity with two successive atoms, considered 
below.

\section*{B. A single cavity and two two-level atoms}

Before considering the real micromaser, we now investigate a 
micromaser-type system where two two-level atoms, the 
first prepared 
in the excited state $|e>$, and the second prepared in the ground 
state $|g>$, are sent into 
a vacuum cavity one after the other, i.e., there is no spatial overlap
between the two atoms. Our purpose here is to demonstrate analytically the
increase of two-atom entanglement in this model with the increase of cavity
damping rate. We compute the time-evolved density state for
the tripartite system of the two atoms and the cavity under the same
secular approximation, and the approximation of a two level (zero or one
photon) cavity.
The reduced density state of the pair of atoms $A_1A_2$ is given by
\begin{eqnarray} 
\rho(t)_{A_1A_2}&=&\textrm{Tr}_{C_1}(\rho_{A_1A_2C_1})
=\left(\begin{array}{ccccc}
\gamma_2 & 0 & 0 & 0\\
0 & \gamma_3 & -\gamma_4 & 0\\
0 & -\gamma_4 & \gamma_1 & 0\\
0 & 0 & 0 & 0  \end{array}\right)
\label{24}
\end{eqnarray}
in the basis of $|g_1g_2>$,
$|g_1e_2>$, $|e_1g_2>$, and $|e_1e_2>$ states, and
where $\gamma_i$ are given by
\begin{eqnarray}
\gamma_1&=&(1-\sin^2{(gt)}e^{-\kappa t}),\nonumber\\
\gamma_2&=&\cos^2{(gt)}\sin^2{(gt)}e^{-2\kappa t},\nonumber\\
\gamma_3&=&\sin^4{(gt)}e^{-2\kappa t},\nonumber\\
\gamma_4&=&\biggl(\sin{(gt)}e^{-\kappa t/2}-\frac{\kappa}{2g}\cos{(gt)}e^{-\kappa t/2}+\frac{\kappa}{2g}\biggr)\times\nonumber\nonumber\\
&&\cos{(gt)}\sin{(gt)}e^{-\kappa t},
\label{24b}
\end{eqnarray}

The concurrence for the joint two-atom state $\rho(t)_{A_1A_2}$ 
is given by
\begin{eqnarray}
{\it C}(\rho(t)_{A_1A_2})&=&2\sin^2{gt}e^{-\kappa t}\sqrt{(1-\sin^2{(gt)}
e^{-\kappa t})}
\label{25}
\end{eqnarray}
For values of $\kappa/g$ and $gt$ such that $(\tan^2{gt})(gt)(\kappa/g) \ll 1$, 
we can write
\begin{eqnarray}
{\it C}(\rho(t)_{A_1A_2}\thickapprox {\it C}_{ideal}(1+\frac{1}{2}\kappa t\tan^2{(gt)}-\kappa t).
\label{26}
\end{eqnarray}
where ${\it C}_{ideal}$ (no dissipation) in this case is given by
${\it C}_{ideal}= 2\cos{gt}\sin^2{gt}$. Thus, enhancement of 
${\it C}(\rho(t)_{A_1A_2})$, i.e., the increase of the atom-atom mixed-state
entanglement over its value in the ideal cavity case is possible
if we choose interaction time judiciously, such that
$\tan{(gt)}>\sqrt{2}$. For fixed $gt$, the concurrence
${\it C}(\rho(t)_{A_1A_2})$ is maximised at 
$\kappa/g=[{\mathrm ln}((3/2)\sin^2{(gt)})]/(gt)$. Further increase  
of cavity damping $\kappa$,
causes the two-atom entanglement to fall off. This sets the stage
for us to investigate next the full micromaser dynamics.

\section*{C. The one-atom micromaser}

We now consider the real micromaser which has been experimentally 
operational\cite{micro1}.  
The mathematical model for the micromaser has been well studied,
and is described in detail in 
\cite{micro3,nayak,englert}. Its essential features are outlined briefly here.
The cavity is pumped to its steady state
by a Poissonian stream of atoms passing through it one at a time, with the 
time of flight through the cavity $t$ being the same for every atom. The 
dynamics of 
these individual flights are governed by the evolution equation with three 
kinds of interactions given by 
\begin{eqnarray}
\dot\rho=\dot\rho|_{{\textrm atom-reservoir}}+\dot\rho|_{{\textrm field-reservoir}}+\dot\rho|_{{\textrm atom-field}},
\label{28}
\end{eqnarray}
where the strength of three couplings are given by the parameters $\Gamma$
(the atomic dissipation constant), $\kappa$ (the cavity leakage constant),
and $g$ (the atom-field interaction constant) with
the individual expressions provided in \cite{micro3,nayak}. 
Obviously, 
$\Gamma=0=g$ describes the dynamics of the cavity when there is no atom inside 
it. The finite temperature of the cavity is represented by the average 
thermal  photons $n_{th}$, obtained from B-E statistics. 

The density matrix of 
the steady-state cavity field $\rho^{ss}_f$ can be obtained by solving the 
above equation and tracing over the reservoir and atomic 
variables\cite{micro3,nayak}. The photon 
distribution function  is then given by 
\begin{eqnarray}
P^{ss}_n=\bra{n}\rho^{ss}_f\ket{n}
\label{29}
\end{eqnarray}
We display the steady state photon statistics for the micromaser
for experimentally realizable values\cite{micro1} of the parameters 
$N$ ($N=R/2\kappa$, where $R$ denotes the number of atoms passing
through the caviy per second), $n_{th}$ and $gt$ in Table~1. The probability 
of getting two ($P_2$) or more photons
inside the cavity is negligible. The photon
statistics thus provides a justification for our earlier assumption of a 
two-level cavity ($P_0$ and $P_1$)
used for obtaining our analytical results of entanglement enhancement
through dissipation in the previous examples. 
However, our present analysis for the
real micromaser does not employ this assumption.

\begin{table}
\begin{center}
\caption{\label{tab:table1}
{\footnotesize Steady state photon statistics for the micromaser with the 
parameter values $n_{th}=0.033$, $N=1$, and $gt=3\pi/4$.}}
\begin{tabular}{|l|l|l|l|l|l|l|}
%\begin{tabular}{|l|l|l|l|l|l|l|}
\hline
$\kappa/g$ & $P_{0}$& $P_{1}$ & $P_{2}$ & $<n>$\\
\hline
0.1&0.771&0.220&0.007&0.236\\
\hline
0.01&0.664&0.316&0.014&0.359\\
\hline
0.005&0.655&0.324 &0.015&0.370\\
\hline
0.0000807&0.645&0.332 &0.016&0.382\\
\hline
\end{tabular}
\end{center}
\end{table}

We compute the atomic entanglement generated between two experimental
atoms that
pass successively through the micromaser cavity.
The tripartite joint state of the cavity and the two atoms 
is obtained by summing over all $n$. The reduced density state of 
the two atoms after passing through the 
the cavity field is given by
\begin{eqnarray} 
\rho(t)_{A_1A_2}&=&\textrm{tr}_{f}
(\rho(t)_{A_1A_2C_1})
= \left(\begin{array}{ccccc}
\beta_5 & 0 & 0 & 0\\
0 & \beta_3 & \beta_4 & 0\\
0 & \beta_4 & \beta_2 & 0\\
0 & 0 & 0 & \beta_1  \end{array}\right)
\label{b31}
\end{eqnarray}
where the $\beta_i$ are given by
\begin{eqnarray}
\beta_1&=&\sum_nP^{ss}_n\cos^4{(\sqrt{n+1}gt)},\nonumber\\
\beta_2&=&\sum_nP^{ss}_n\cos^2{(\sqrt{n+1}gt)}
\sin^2{(\sqrt{n+1}gt)},\nonumber\\
\beta_3&=&\sum_nP^{ss}_n\cos^2{(\sqrt{n+2}gt)}\sin^2{(\sqrt{n+1}gt)},\nonumber\\
\beta_4&=&\sum_nP^{ss}_n\sin^2{(\sqrt{n+1}gt)}\cos{(\sqrt{n+1}gt)}\cos{(\sqrt{n+2}gt)},\nonumber\\
\beta_5&=&\sum_nP^{ss}_n\sin^2{(\sqrt{n+1}gt)}\sin^2{(\sqrt{n+2}gt)}.
\label{b32}
\end{eqnarray}

\vskip 0.4cm
\begin{figure}[h]
	\begin{center}
	\includegraphics[width=8cm]{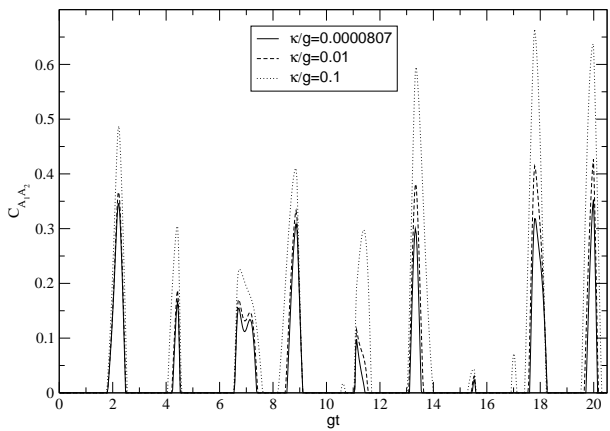}
         \caption{The two-atom concurrence $C_{A_1A_2}$ is plotted 
versus the Rabi angle $gt$ for various values of the cavity
leakage parameter $\kappa$. Here we choose $N=1$ and  $n_{th}=0.033$}
        \end{center}
         \end{figure}

The  concurrence of the two-atom state is plotted 
with respect to the Rabi angle
$gt$ in Fig. 1 choosing the cavity temperature as in 
the operational micromaser\cite{micro1}. 
We see that the entanglement between the two atoms increases 
as we increase the cavity 
dissipation parameter $\kappa/g$ from the experimental values (solid curves
in Fig. 1) at a fixed values of the Rabi angle $gt$. 
Increased damping of the micromaser cavity causes the average cavity
photon number $<n>$ to go down, as displayed in Table~1. The collective 
dynamics of the system causes the magnitude of two-atom entanglement 
to rise with decreasing $<n>$. This anti-correlation of the two-atom
entanglement with the cavity photon number $<n>$ has also been observed
in earlier works\cite{bip}. It is expected that further increase of $\kappa/g$
beyond the values shown in the figure would make the concurrence to fall.
However, the validity of the micromaser theory that we have used\cite{nayak} 
is itself limited to low dissipation values by the secular 
approximation\cite{secular}. 

To summarize, in this Letter we have presented concrete examples of
the increase of entanglement caused by the stronger interaction of
a part of a compsite system with its environment. We have first considered
the atom-cavity entanglement in a system comprised of two entangled 
cavities and a two-level atom\cite{David}. The derived expression for
the atom-cavity concurrence clearly shows the maximization of entanglement
for intermediate values of the cavity damping. A similar analytical result
is also obtained for two-atom entanglement in the case of a 
micromaser-type system involving a single cavity and two atoms. We have 
further investigated a model for the real micromaser\cite{micro1}
and demonstrated the increase of atomic entanglement with cavity damping
for fixed atom-cavity interaction times
for experimentally operational values of the micromaser parameters.
With further development, it may be possible to utlize this effect of
environment assisted entanglement enhancement in information processing
involving multipartite systems where the interactions times may
not be easily controllable. 
In conclusion, we highlight that atom-cavity systems provide much
scope for the quantitative tests of several manifestations of environment 
induced entanglement\cite{plenio,braun,knight} in experimentally
realizable situations. 
 
We are grateful to G. S. Agrawal for his comments on the
work which shaped the paper.

%%%%%%%%%%%%%%%%%%%%%%%%%%%%%%%%%%%%%%%%%%%%%%%%%%%%%%%%%%%%%%%%%%

%%%%%%%%%%%%%%%%%%%%%%%%%%%%%%%%%%%%%%%%%%%%%%%%%%%%%%%%%%%%%                                                                       

\begin{thebibliography}{99}

\bibitem{Nielsen}
M. A. Nielsen  and I. L. Chuang, {\it Quantum Computation and 
Quantum Information} (Cambridge University Press, Cambridge, UK 2000).

\bibitem{proposals}
See, for example, 
J. I. Cirac and P. Zoller, Phys. Rev. A{\bf 50}, R2799 (1994);
J. I. Cirac, P. Zoller, H. J. Kimble, and H. Mabuchi,
Phys. Rev. Lett. {\bf 78}, 3221 (1997);
L. -M. Duan, M. D. Lukin, 
J. I. Cirac and P. Zoller, Nature, 
{\bf 414}, 413 (2001); E. Solano, G. S. Agarwal,  and H. Walther, Phys. Rev. 
Lett. {\bf 90} 027903 (2003); S. G. Clark and A. S. 
Parkins, Phys. Rev. Lett. {\bf 90}, 047905 (2003); L.-M. Duan, B. Wang, 
H. J. Kimble, Phys. Rev. A{\bf 72}, 032333 (2005); G. S. Agarwal and K. T.
Kapale, Phys. Rev. A{\bf 73}, 022315 (2006).

\bibitem{examples}
See, for instance, E. Hagley, X. Maître, G. Nogues, C. Wunderlich, M. Brune, 
J. M. Raimond, 
and S. Haroche, Phys. Rev. Lett. {\bf 79}, 1 (1997);
A. Rauschenbeutel, P. Bertet, S. Osnaghi, G. Nogues, M. 
Brune, J. M. Raimond and S. Haroche, Phys. Rev. A {\bf 64}, 050301(R) (2001);
A. Auffeves, P. Maioli, T. Meunier, S. Gleyzes, G. Nogues, M. Brune, 
J. M. Raimond, and S. Haroche, Phys. Rev. Lett. {\bf 91}, 230405 (2003);
S. Nußmann, M. Hijlkema, B. Weber, F. Rohde, G. Rempe, and A. Kuhn
Phy. Rev. Lett. {\bf 95}, 173602 (2005);
D. N. Matsukevich, T. Chanelière, S. D. Jenkins, S.-Y. Lan, T. A. B. Kennedy,
and A. Kuzmich, Phys. Rev. Lett. {\bf 96}, 030405 (2006).

\bibitem{Raimond} 
For a review, see, J. M. Raimond, M. Brune and S. Haroche, 
Rev. Mod. Phys. {\bf 73}, 565 (2001).

\bibitem{zurek}
W. H. Zurek, Phys. Rev. D{\bf 26}, 1862 (1982); E. Joos and H. D. Zeh, Z. Phys.
B{\bf 59}, 223 (1985).

\bibitem{plenio}
M. B. Plenio, S. F. Huelga, A. Beige, and P. L. Knight,
Phys. Rev. A{\bf 59}, 2468 (1999); S. Bose, P. L. Knight, M. B. Plenio, and 
V. Vedral, Phys. Rev. Lett. {\bf 83}, 5158 (1999); A. Beige, S. Bose, D. Braun,
S. F. Hulega, P. L. Knight, M. B. Plenio and V. Vedral, J. Mod. Opt. {\bf 47},
2583 (2000); L. M. Duan and H. J. Kimble, Phys. Rev. Lett. {\bf 90}, 253601
(2003).

\bibitem{braun}
D. Braun, Phys. Rev. Lett. {\bf 89}, 277901 (2002).

\bibitem{knight}
M. Kim, J. Lee, D. Ahn and P. L. Knight, Phys. Rev. A{\bf 65}, 040101(R) (2002);
B. Ghosh, A. S. Majumdar and N. Nayak, quant-ph/0603039 (to appear in IJQI).

\bibitem{others}
S. Schneider and G. J. Milburn, Phys. Rev. A{\bf 65}, 042107 (2002); B. Kraus
and J. I. Cirac, Phys. Rev. Lett. {\bf 92}, 013602 (2004); S.-B. Li and J.-B.
Xu, quant-ph/0505216.

\bibitem{plenio2}
M. B. Plenio and S. F. Hulega, Phys. Rev. Lett. {\bf 88}, 197901 (2002).

\bibitem{book}
G. S. Agarwal, in {\it Springer Tracts in Modern Physics}, {\bf 70}, Ed.: G. 
H\"ohler, (Springer, 
New York and Berlin 1974).  
\bibitem{micro1}
G. Rempe, F. Schmidt-Kaler and H. Walther, Phys. Rev. Lett. 
{\bf 64}, 2783 (1990); M. Brune, E. Hagley, J. Dreyer, X. Maître, A. Maali, 
C. Wunderlich, J. M. Raimond, and S. Haroche, Phys. Rev. Lett. {\bf 77}, 4887 
(1996); E. Hagley, X. Maître, G. Nogues, C. Wunderlich, 
M. Brune, J. M. Raimond, and S. Haroche, Phys. Rev. Lett. {\bf 79}, 1 (1997);
M. Weidinger, B. T. H. Varcoe, R. Heerlein, and H. Walther,
Phys. Rev. Lett. {\bf 82}, 3795 (1999).

\bibitem{micro2}
S. J. D. Phoenix and S. M. Barnett, J. Mod. Opt. 40, 979 (1993);
J. I. Cirac and P. Zoller, Phys. Rev. A{\bf 50}, R2799 (1994).

\bibitem{micro3}
A. S. Majumdar and N. Nayak, Phys. Rev. A {\bf 64}, 013821 (2001);
A. Datta, B. Ghosh, A. S. Majumdar and N. Nayak,
Europhys. Lett. 67, 934 (2004). 

\bibitem{David}
L. Davidovich, N. Zagury, M. Brune, J. M. Raimond and S. Haroche, Phys. Rev. 
A{\bf 50} R895 (1994).

\bibitem{secular}
S. Haroche and J. M. Raimond, in ``Advances in atomic and molecular physics'',
Vol. 20, eds. D. R. Bates and B. Bederson (Academic, New York, 1985).

\bibitem{con}S. Hill and W. K. Wootters, Phys. Rev. Lett. {\bf78}, 5022
(1997); W. K. Wootters, Phys. Rev. Lett. {\bf80}, 2245 (1998).

\bibitem{ckw}
V. Coffman, J. Kundu and W. K. Wootters, Phys. Rev. A {\bf 61},
052306 (2000).

\bibitem{monogam2}
B. Ghosh, A. S. Majumdar and N. Nayak,
Int. J. Quant. Inf. {\bf 4}, 665 (2006).

\bibitem{nayak}
N. Nayak, Opt. Commun. {\bf 118}, 114 (1995).

\bibitem{englert}
B. G. Englert, P. Lougovski, E. Solano and H. Walther, 
Laser Phys. {\bf 13}, 355 (2003).

\bibitem{bip}
B. Ghosh, A. S. Majumdar and N. Nayak, quant-ph/0603039 
(to appear in Int. J. Quant. Inf.);  B. Ghosh, A. S. Majumdar and N. Nayak, 
quant-ph/0605191.

\end{thebibliography}
\end{document}